\documentclass[9pt,twocolumn,twoside]{revtex4-1}

\usepackage{amsmath}
\usepackage{graphicx}          
\usepackage{dcolumn}
\usepackage{amssymb,amsmath,bm}
\usepackage{color}
\usepackage{epstopdf}
\usepackage{caption}
\usepackage{subcaption}
\usepackage{float} 
\usepackage{anyfontsize}

\usepackage{subcaption}
\usepackage{ragged2e}
\DeclareCaptionJustification{justified}{\justifying}
\newcommand{\ba}{\begin{array}}
\newcommand{\ea}{\end{array}}
\newcommand{\be}{\begin{equation}}
\newcommand{\ee}{\end{equation}}
\newcommand{\bea}{\begin{eqnarray}}
\newcommand{\eea}{\end{eqnarray}}

\begin{document}
\title{Vortex coronagraphy from self-engineered liquid crystal spin-orbit masks}

\author{Artur Aleksanyan and Etienne Brasselet*}
\affiliation{Univ. Bordeaux, LOMA, UMR 5798, F-33400 Talence, France}
\affiliation{CNRS, LOMA, UMR 5798, F-33400 Talence, France}
\affiliation{Corresponding author: etienne.brasselet@u-bordeaux.fr}

\begin{abstract}
We report on a soft route toward optical vortex coronagraphy based on self-engineered electrically tunable vortex masks made of liquid crystal topological defects. These results suggest that a nature-assisted technological approach to the fabrication of complex phase masks could be useful in optical imaging whenever optical phase singularities are at play.
\end{abstract}

\maketitle


The observation of faint objects near a bright source of light is a basic challenge of high-contrast imaging techniques such as the quest for extrasolar planets in astronomical imaging. This has led to the development of instruments called coronagraphs, which combine the high extinction of stellar light and the high transmission of a low-level signal at small angular separation. Stellar coronagraphy was initiated more than 80 years ago when Lyot studied solar corona without an eclipse by selective occultation of sunlight, placing an opaque disk in the focal plane of a telescope \cite{lyot_mnras_1939}. On the other hand, phase mask coronagraphs offer good performance for the observation of point-like sources. An early version consisted of using a disk $\pi$-phase mask \cite{roddier_pasp_1997} whose chromatic drawback arising from discrete radial phase step was solved a few years later by incorporating discrete radial \cite{soummer_aa_2003} or azimuthal \cite{rouan_pasp_2000} phase modulation to the original design. Later on, continuous azimuthal phase ramps improved the approach \cite{mawet_apj_2005, foo_ol_2005} and led to the advent of optical vortex coronagraphy.\raggedbottom

Vortex coronagraphs rely on the selective peripheral redistribution of on-axis  light outside an area of null intensity at the exit pupil plane of the instrument, which is done by placing a spiraling phase mask in the Fourier plane characterized by a complex transmittance of the form $\exp(i\ell \phi)$, where the charge $\ell$ is an even integer and $\phi$ is the usual azimuthal angle in the transverse plane. This enables optimal on-axis rejection of light by placing an iris (called Lyot stop) at the exit pupil plane, while the off-axis weak signal is almost unaffected for an angular separation larger than the diffraction limit \cite{swartzlander_jo_2009}. There are two families of optical vortex phase masks that rely on the scalar (phase) and vectorial (polarization degree of freedom) properties of light. The former case refers to the helical shaping of the wavefront from a refractive phase mask, whereas the latter one exploits the polarization properties of space-variant birefringent optical elements. Indeed, inhomogeneous anisotropic phase masks endowed with azimuthal optical axis orientation of the form $\psi(\phi) = m\phi$ with $m$ half-integer and associated with half-wave birefringent phase retardation lead to a vortex mask of charge $\ell= 2\sigma m$ for incident on-axis circularly polarized light beam with helicity $\sigma = \pm 1$, as originally shown in \cite{biener_ol_2002} using form birefringence  (subwavelength grating of dielectric materials) and, in \cite{marrucci_prl_2006}, using true birefringence.

The achromatic features of the vectorial option versus its scalar counterpart \cite{mawet_oe_2009, murakami_oe_2013} have eventually led to equip state-of-the-art large instruments such as
Keck
, Subaru
, Hale
, Large Binocular Telescope, 
and Very Large Telescope 
with vectorial vortex coronagraphs. In practice, all these installations exploit vortex masks obtained either from brute force nanofabrication, where form birefringence arises from subwavelength material structuring \cite{biener_ol_2002}, or liquid crystal polymers technology, where true birefringence is patterned on demand \cite{mceldowney_ol_2008}. Still, these high-tech "writing" processes ultimately suffer from fabrication constraints and finite resolution,  preventing the creation of ideal material singularity, which implies various trade-offs. One can mention the preferential use of natural, rather than a form-birefringence option at shorter wavelengths \cite{mawet_oe_2009}, and the use of an opaque disk to reduce the detrimental influence of central misorientation of the engineered optical axis pattern \cite{mawet_apj_2010, nersisyan_oe_2013}. The realization of higher-order masks with even topological charges $\ell > 2$ \cite{mawet_apj_2005} that are desirable for future extremely large telescopes \cite{delacroix_spie_2014} is another manufacturing challenge. 
Since the basic technological bottleneck can be identified as the man-made technology itself, a nature-assisted approach would likely open a novel generation of vortex masks.

Various kinds of spontaneously formed nematic liquid crystal topological defects have been previously demonstrated to behave as natural \cite{brasselet_prl_2009, loussert_prl_2013} or field-induced \cite{brasselet_pra_2010, brasselet_ol_2011, barboza_prl_2012} vectorial optical vortex generators with even charge $|\ell|=2$ and optimal beam shaping characteristics. Optical vortex masks realized without the need for a machining technique have also been obtained from other mesophases such as cholesteric \cite{yang_jo_2013} and smectic \cite{son_oe_2014} liquid crystals, but at the expense of efficiency.  Here, we propose exploring the potential of self-engineered liquid crystal vortex phase masks for high-contrast imaging applications, in particular, in the case of optical vortex coronagraphy.

We choose the so-called umbilical defects that appear in homeotropic nematic liquid crystal films with negative dielectric anisotropy under a quasi-static electric field and above a threshold voltage $U=U^*$ of a few volts \cite{rapini_jp_1973}. These defects are associated with optical axis orientation angle of the form $\psi(\phi) = m\phi + \psi_0$ with $m=\pm1$ and $\psi_0$ a constant; see Figs.~1(a) and 1(b) for imaging between crossed linear polarizers. It has been shown previously that such defects behave as spin-orbit optical vortex generators \cite{brasselet_ol_2011}. This can be described by neglecting the diffraction inside the optical element itself given the circularly polarized incident field in the plane of the sample, $E(r)\,{\bf c}_\sigma$, where $r=0$ refers to the defect location and ${\bf c}_\sigma = ({\bf x} + i\sigma {\bf y})/\sqrt{2}$ refers to circular polarization basis. In the case of transparent media as is our case, the output light field can be expressed as \cite{brasselet_olcomm_2013}
\bea
\label{Eout}
\nonumber &&\hspace{-10mm}{\bf E}_{\rm out} (r,\phi) \propto E(r) \exp[i\Delta(r)/2] \big\{ \cos[\Delta(r)/2]\, {\bf c}_\sigma\\ &&\hspace{15mm}+ i \sin[\Delta(r)/2] \exp[i2\sigma \psi(\phi) ] \,{\bf c}_{-\sigma}\big\}\,,
\eea
where $\Delta(r)$ is the $r$-dependent birefringent phase retardation modeled following the work of Rapini \cite{rapini_jp_1973}. Within such a description, one has $\Delta(r) = \delta(r/r_{\rm c}) \Delta_\infty$, where $r_{\rm c} = (L/\pi) ({\cal K}/K_3)^{1/2} [(U/U^*)^2 - 1]^{-1/2}$ is the voltage-dependent defect core radius with $L$ being
the nematic film thickness; ${\cal K} = K_2$ for $m=+1$ and $\psi_0=\pi/2$ (our case) and ${\cal K} = (K_1+K_2)/2$ for $m=-1$ whatever $\psi_0$, with $K_{i=1,2,3}$ the Frank elastic constant of splay, twist, and bend director distortions \cite{oswald_book_nematic}; and $\Delta_\infty$ is the voltage-dependent asymptotic value of $\Delta$ at large $r$. The calculated voltage-independent function $\delta(r/r_{\rm c})$, which describes how the retardance depends on the distance to the defect core, and reduced core radius $r_{\rm c}/L$ versus reduced voltage $U/U^*$ are shown in Figs.~1(c) and 1(d).
\begin{figure}[b!]
\centering\includegraphics[width=1\columnwidth]{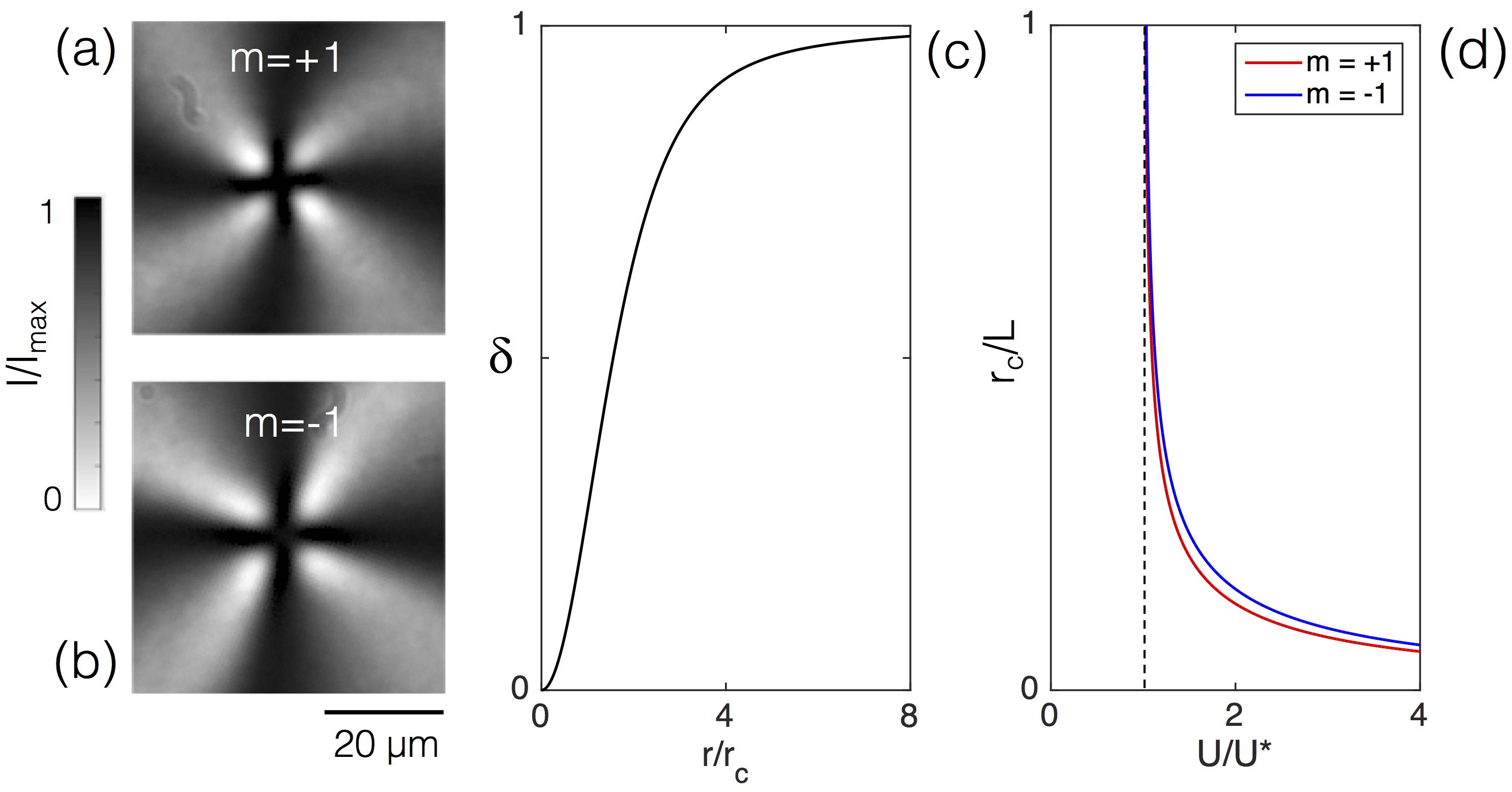}
\caption{
(a,b) Typical images of nematic umbilical defects with a topological charge $m=\pm1$ observed between crossed linear polarizers.
(c) Calculated radial dependence of the function $\delta$ from Rapini's model. (d) Calculated reduced core radius $r_{\rm c}/L$ versus reduced voltage $U/U^*$ for $m=\pm1$ (red/blue curves).
}
\end{figure}

From Eq.~(\ref{Eout}), the spin-orbit generation of a spiraling optical phase with charge $\pm2$ for the helicity-flipped output field component is never ideal due to the space-variant retardance, which is quantitatively evaluated by the purity
\be
\label{eta}
\eta = \int_0^\infty |E(r)|^2 \sin^2[\Delta(r)/2] r dr \Big/\!\! \int_0^\infty |E(r)|^2 r dr \,,
\ee
$0 < \eta < 1$, while the spin-orbit mask is described up to an unimportant phase factor by the complex amplitude transmittance
\be
\label{tau_umbilic}
\tau(r,\phi) = \sin[\Delta(r)/2] \exp[i\Delta(r)/2+2i\sigma \psi(\phi)]\,.
\ee

Looking for a complex transmittance of the form $\exp(i\ell\phi)$ one should ideally target an umbilic with $\Delta_\infty=\pi$ (that defines the voltage $U=U_\pi$) and core radius $r_c$ smaller  than the characteristic beam spot size $w$ in the plane of the sample. 
Here, we choose a $L=10~\mu$m-thick nematic film made from a glass cell (from EHC Ltd) whose inner surface is provided with transparent electrodes. 
\begin{figure}[b!]
\centering\includegraphics[width=1\columnwidth]{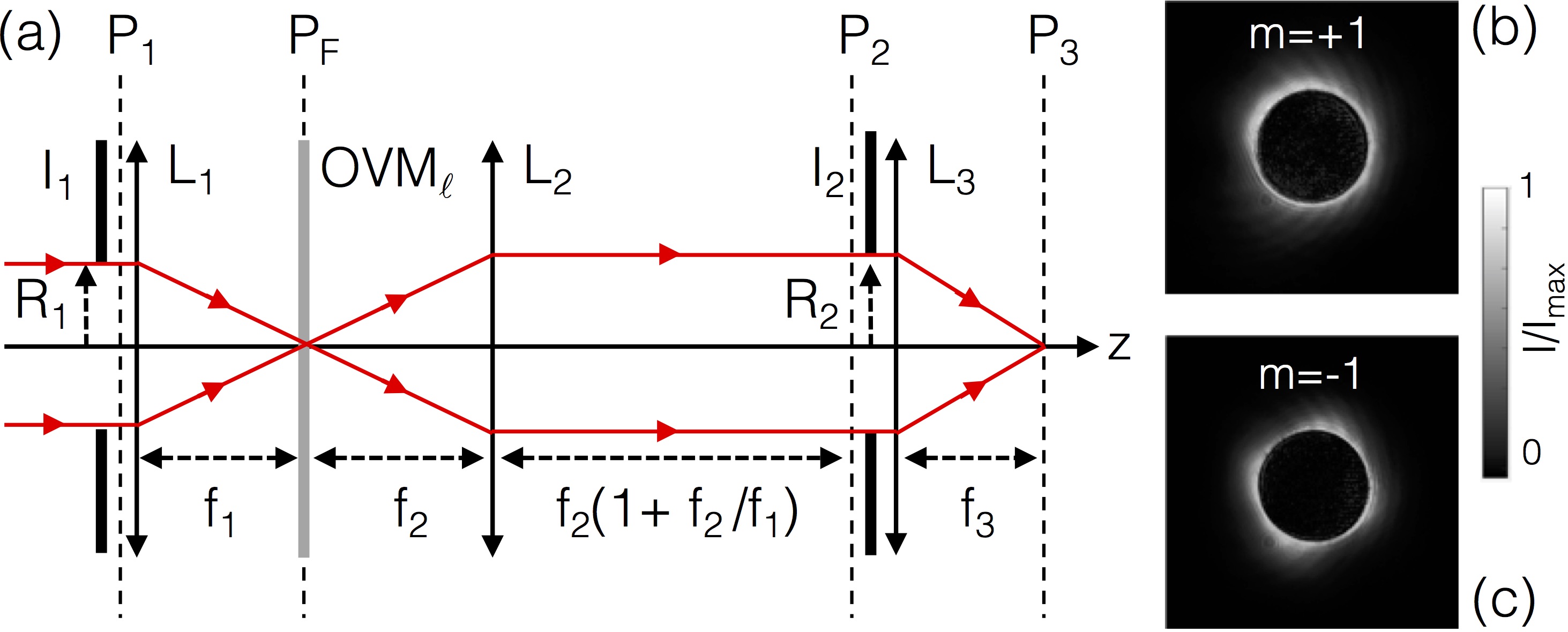}
\caption{
(a) Optical setup. It consists of a set of irises $I_{1,2}$ and lenses $L_{1,2,3}$ with focal lengths $f_{1,2,3}$, and an optical vortex mask of the topological charge $\ell$, OVM$_\ell$, placed in the Fourier plane P$_{\rm F}$ of the telescope formed by lenses $L_{1,2}$. The planes $P_{1,2}$ refer to the entrance and exit pupil planes, respectively, while the plane $P_{3}$ refers to the imaging plane. The red lines refer to the ray optics tracing. (b) and (c) Rings of fire observed in the exit pupil plane $P_2$ for $m=\pm1$.
}
\end{figure}
The liquid crystal film is prepared from the nematic MLC-2079 (from Licristal) having a dielectric relative permittivity of $\varepsilon_\parallel = 4.1$ along the molecular alignment and $\varepsilon_\perp = 10.2$ perpendicular to it (tabulated data at 1~kHz frequency) and refractive indices $n_\parallel=1.64$ and $n_\perp=1.49$ at a 589~nm wavelength. The Frank elastic constant of splay, twist, and bend director distortions \cite{oswald_book_nematic} are  $K_1=15.9$~pN (tabulated), $K_2=9.5$~pN (measured) and $K_3=18.3$~pN (tabulated), respectively. This gives $U^*= 1.83$~V$_{\rm rms}$, $U_\pi=2.52$~V$_{\rm rms}$\cite{rapini_jp_1973} and $r_{\rm c} = (2.4,2.8)~\mu$m for $m=\pm 1$.

In practice, once $U$ is set above $U^*$, a few defects remain at a fixed location in the sample after the annihilation dynamics between nearby topological defects with the opposite topological charges has taken place. Typically, experiments are performed a couple of hours or more after the voltage, whose steady value is set at $U_\pi$, has been switched on. Then, a defect is placed on-axis in the focal plane (P$_{\rm F}$) of a telescope, as depicted in Fig.~2(a), which sketches the main part of the optical arrangement used to simulate a vortex coronagraph in the laboratory where starlight is mimicked by an on-axis incident quasi-plane wave. Experiments are performed using point-like sources generated from He-Ne lasers  at wavelength $\lambda = 633$~nm and located in the focal plane of the microscope objectives ($10\times$, NA$=0.25$) with an overfilled entrance pupil. The obtained light beams are collimated by lenses with a 20~cm focal length, apertured by an iris with 25~mm diameter, and remotely redirected at an angle $\alpha$ from the optical axis $z$ by mirrors. 
\begin{figure}[b!]
\centering\includegraphics[width=1\columnwidth]{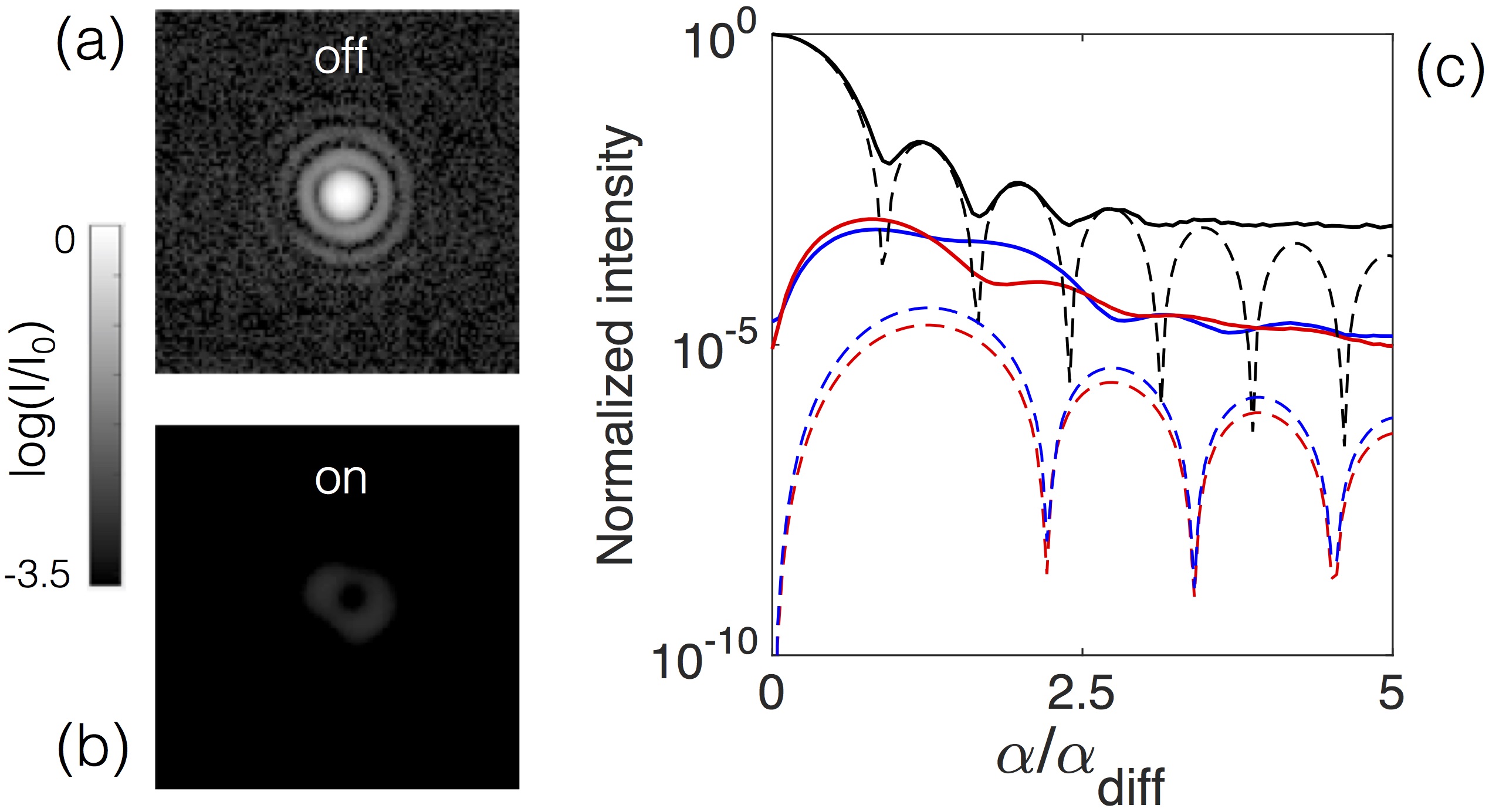}
\caption{
(a) and (b) Typical on-axis point-like source images in the plane P$_3$ when the coronagraph is off/on, $I_0$: maximal intensity in the off state, while the intensity range refers to the 4096 camera levels. (c) Azimuth-averaged angular intensity profile in off (black curves) and on (color curves) states. Solid curves, experimental data; dashed curves, simulations; red curves, $m=+1$; blue curves, $m=-1$.
}
\end{figure}
Lenses L$_1$ and L$_2$ [see Fig.~2(a)] are identical microscope objectives ($4\times$, NA $=0.1$, entrance pupil diameter $R=4.42$~mm); the radii of I$_1$ and I$_2$ are $R_1=1$~mm and $R_2=0.75$~mm, respectively. This gives $w=0.61\lambda/(0.1R_1/R)=17~\mu$m. The imaging of the sources is made by lens L$_3$ with a focal length $f_3=30$~cm. The input circular polarization is ensured by the polarizer and quarter-wave plate, whereas the contra-circularly polarized field component after the optical vortex mask OVM is selected by another set of quarter-wave plates and polarizers (not shown in Fig.~2). Finally, all images are recorded by 12-bit room-temperature camera.

The coronagraphic behavior of our self-engineered spin-orbit masks is illustrated in Figs.~2(b) and 2(c) which show the so-called "rings of fire" surrounding a dark area with a diameter $2R_1f_2/f_1$ (hence $2R_1$ in our case) in the exit pupil plane P$_2$. The four-fold rotational symmetry for $m =-1$ is reminiscent of the elastic anisotropy of the liquid crystal. Indeed, while only the twist distortions are involved for $m=+1$ when $K_2<K_1$, both chiral and splay distortions at play for $m=-1$ \cite{rapini_jp_1973} break the axisymmetry; see \cite{clerc_pre_2014} for recent numerical investigations. Ensuing starlight rejection is demonstrated in Figs.~3(a) and 3(b) which refer to the image of the on-axis source that is collected in the plane P$_3$ when the coronagraph is "on" (i.e., presence of Lyot stop I$_2$) or "off" (i.e., I$_2$ is removed) while the azimuth-averaged radial intensity profiles in the plane P$_2$ are shown as solid curves in Fig.~3(c).
\begin{figure}[t!]
\centering\includegraphics[width=1\columnwidth]{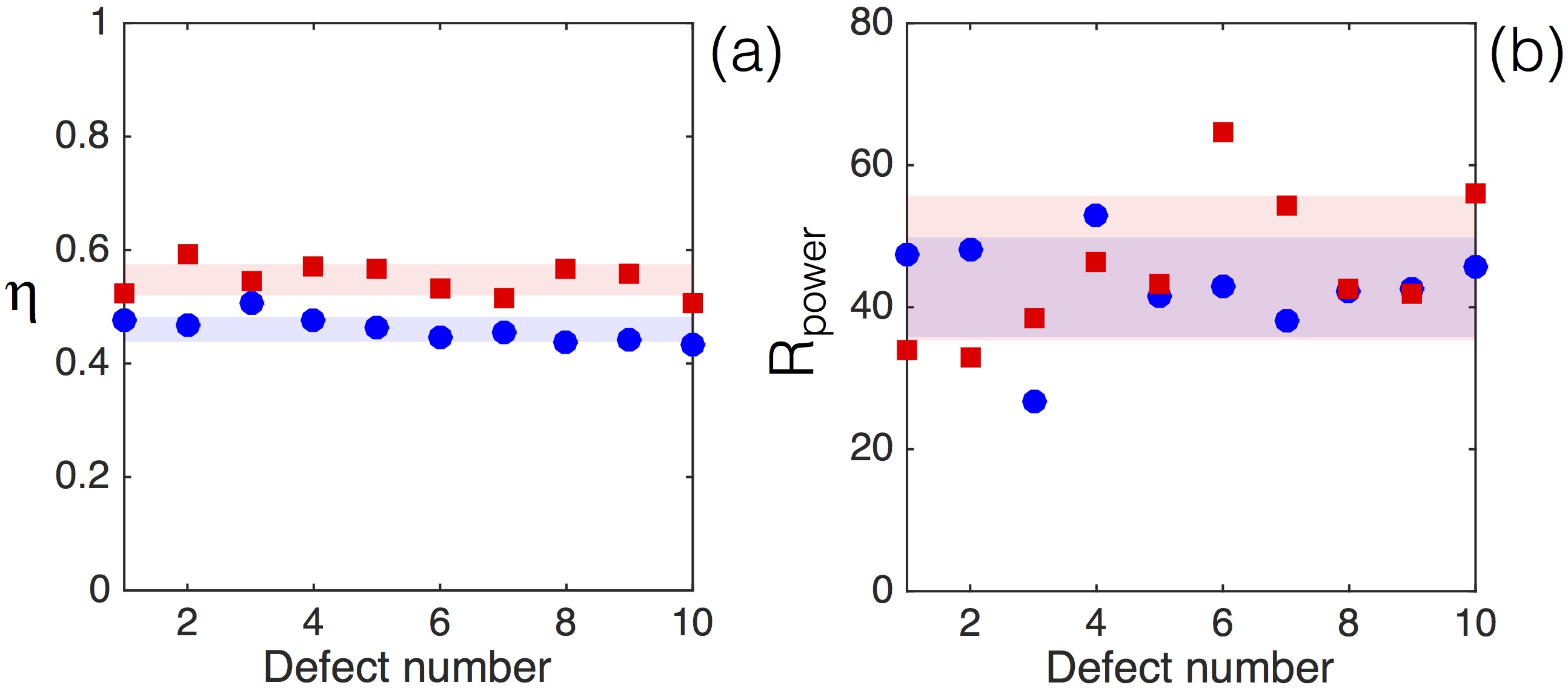}
\caption{
Spin-orbit masks performances statistics. (a) Purity $\eta$ of the optical vortex generation process. (b) Power rejection rate ${\cal R}_{\rm power}$. The colored area refer to standard deviation region. Red square markers, $m=+1$; blue circle markers, $m=-1$. 
}
\end{figure}
\begin{figure}[b!]
\centering\includegraphics[width=1\columnwidth]{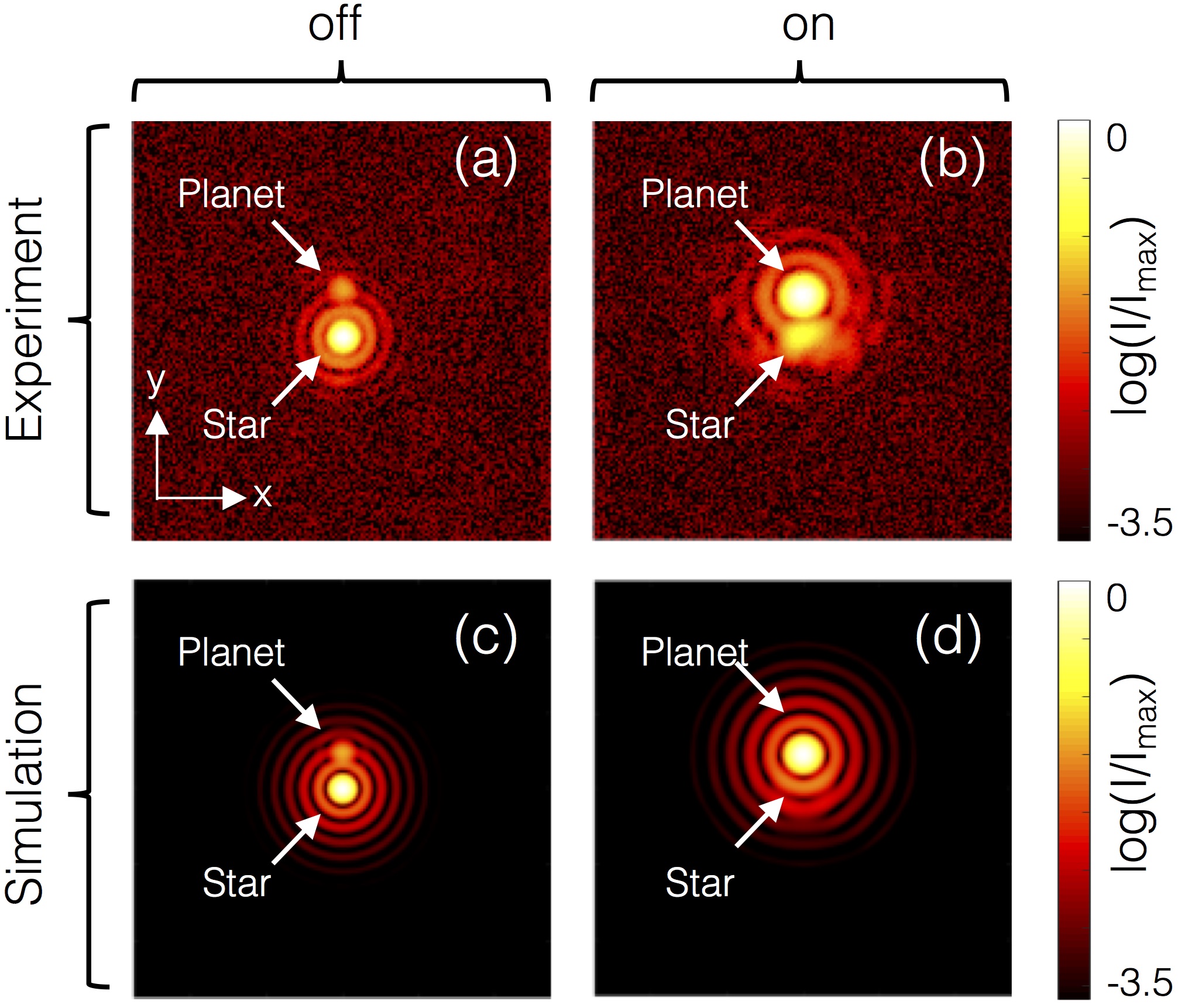}
\caption{
Imaging a star/planet system when the coronagraph is off (without the Lyot stop, which gives a smaller point spread function than the case when iris I$_2$ is present) and on. (a) and (b) Experimental data. (c) and (d) Simulations.
}
\end{figure}
Quantitatively, the coronagraph performances are gauged from the rejection of on-axis illumination. A peak-to-peak starlight rejection ratio of ${\cal R}_{\rm peak,\, exp} \approx 1000$ is obtained and is associated with a typical power rejection rate ${\cal R}_{\rm power,\, exp} \approx 50$ defined as the ratio between the power outside and inside the Lyot stop:
\be
\label{rejection}
{\cal R}_{\rm power} = \int_{R_2}^\infty I_{\rm 2}(r) r dr \Big/\!\! \int_0^{R_2} I_{\rm 2}(r) r dr \,.
\ee
where $I_{\rm 2}(r)$ is the intensity profile in the plane P$_2$. Statistics from ten defects for $m=\pm1$ is summarized in Fig.~4. The purity $\eta$ shown in Fig.~4(a) is evaluated as the ratio between the power of the generated vortex beam (post-selection circular polarizer is present) and the total output power (post-selection circular polarizer is removed). The experimental data are $\eta_{\rm exp} \simeq (0.55\pm 0.02,0.46\pm 0.02)$ for $m=\pm 1$. Statistics of measured power rejection rates are shown in Fig.~4(b) which gives ${\cal R}_{\rm power,\, exp} \simeq (46 \pm 9, 43 \pm 7)$ for $m=\pm 1$.

Simulations are carried out under the assumption of a paraxial on-axis plane wave impinging on the iris I$_1$, namely by considering $E_1 \propto {\rm circ}(r/R_1)$ as the entrance pupil field in the plane P$_1$, where ${\rm circ}(r/\rho)=0$ for $r>\rho$ and ${\rm circ}(r/\rho)=1$ for $r<\rho$. Then, $\eta$ is calculated by inserting the expression of starlight intensity profile in the plane P$_{\rm F}$, namely $I_{\rm F} = |{\cal F}[E_1]|^2$, in Eq.~(\ref{eta}) where ${\cal F}$ refers to the Fourier transform. This gives $\eta_{\rm model}=(0.80,0.75)$ for $m=\pm 1$. On the other hand $I_{\rm 2}(r) = |{\cal F}^{-1}[\tau(r,\phi){\cal F}[E_1]]|^2$, where ${\cal F}^{-1}$ refers to the Fourier transform, and gives ${\cal R}_{\rm peak, \,model} \simeq (5 \times 10^4,2\times 10^4)$ for $m=\pm 1$. Finally, inserting the latter expression of $I_{\rm 2}(r)$ in Eq.~(\ref{rejection}) gives ${\cal R}_{\rm power, \,model} \approx (3000,1500)$ for $m=\pm 1$. 
As usual, in vortex coronagraphy, simulations predict better performances than experiment. Indeed, any source of material and optical imperfections is not taken into account. Still, our experimental values are rather promising for a first attempt, noting that early \cite{mawet_oe_2009,serabyn_nature_2010} and recent \cite{delacroix_spie_2016} laboratory and on-sky astronomical observations with an artificial vectorial vortex mask reported a peak-to-peak stellar rejection ratio $<$100.

Next, we implement the coronagraphic observation of a "star/planet" system. This is done by adding a faint off-axis illumination (the "planet") at an angle $2\alpha_{\rm diff}$ ($\alpha_{\rm diff}=0.61\lambda/R_1$ is the diffraction limit) from the on-axis light (the "star") obtained from a distinct laser source. The experimental situation without coronagraph is shown in Fig.~5(a) where the peak-to-peak intensity ratio between the star and the planet in the imaging plane P$_3$ is $I_{\rm star}/I_{\rm planet} = 20$ for the purpose of demonstration. The latter ratio is drastically reduced to 0.15 when the vortex mask is electrically turned on; see Fig.~5(b).  Data are compared with simulations taking the same conditions as in experiments.
\begin{figure}[t!]
\centering\includegraphics[width=1\columnwidth]{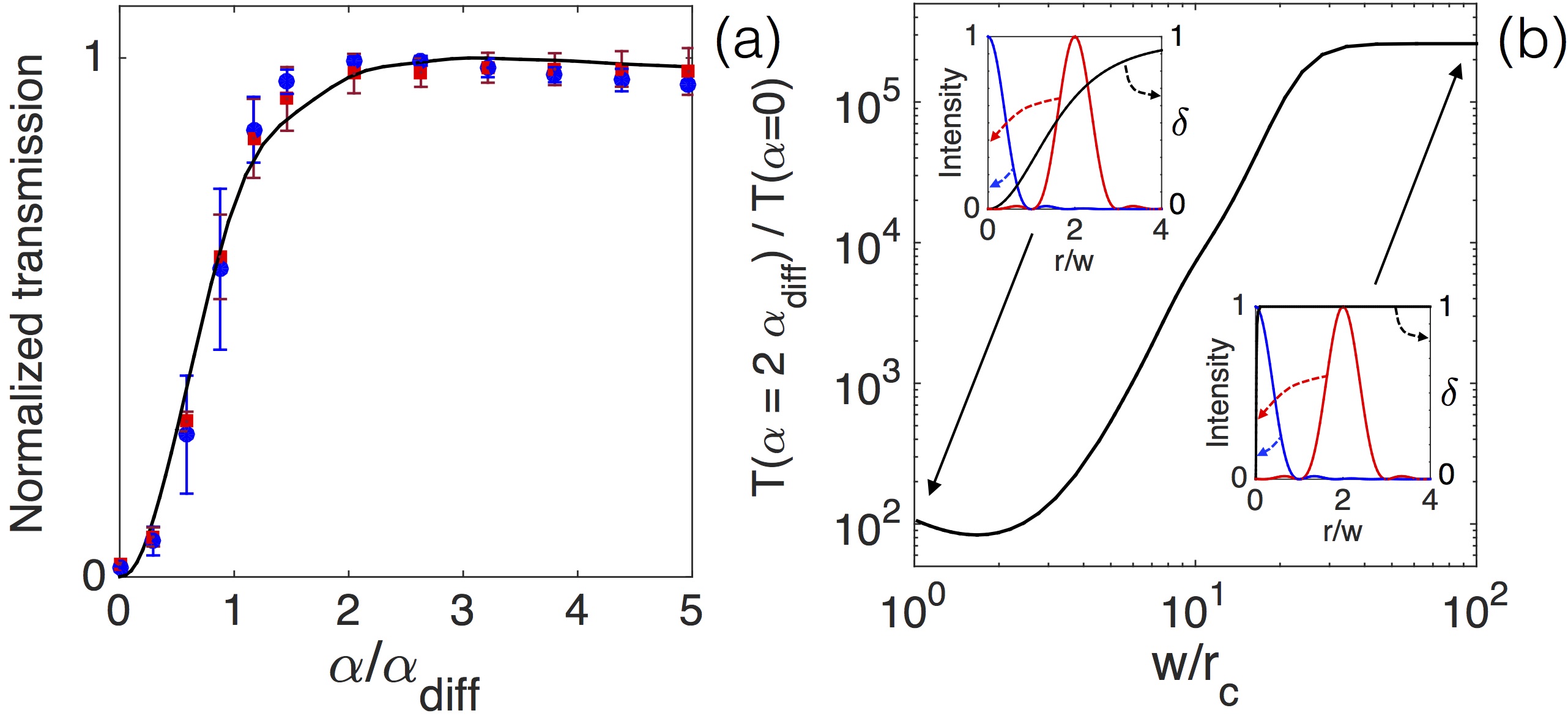}
\caption{
(a) Transmitted power ratio $P_{\rm on} / P_{\rm off}$ as a function of the normalized angular separation distance $\alpha / \alpha_{\rm diff}$ to the telescope axis for $m=\pm1$. Markers, measurements with red/blue color for $m=\pm1$. Solid curves, simulations. (b) Calculated ratio $T(2\alpha_{\rm diff})/T(0)$. Insets, normalized intensity profiles of on-axis ($\alpha=0$) and off-axis ($\alpha=2\alpha_{\rm diff}$) point sources and $\delta$ in the plane P$_{\rm F}$ for $w/r_{\rm c} = 1$ and $w/r_{\rm c} = 100$.
}
\end{figure}
Namely, the images of the star and the planet are calculated from $I_j = |{\cal F}[E_j]|^2$ when the coronagraph is off (Lyot stop is removed) and $I_j = |{\cal F}[{\rm circ}(r/R_2){\cal F}^{-1}[\tau(r,\phi){\cal F}[E_j]]]|^2$ when the coronagraph is on, with $E_{\rm star}=\sqrt{20}\,{\rm circ}(r/R_1)$ and $E_{\rm planet}={\rm circ}(r/R_1)\exp(i2k\alpha_{\rm diff}y)$, $k=2\pi/\lambda$. 
More quantitatively, the coronagraphic throughput is experimentally assessed from the transmitted power ratio $P_{\rm on} / P_{\rm off}$ of the light field passing through the Lyot stop when the coronagraph is on and off as a function of $\alpha / \alpha_{\rm diff}$. Figure ~6(a) summarizes the data obtained for three independent defects for each $m$, where the markers refer to the experiment while the solid curve corresponds to simulations, which exhibit good agreement. Moreover, the central apodization effect of the spin-orbit vortex mask due to the $r$-dependent retardance profile (Fig.~1(a)) is illustrated in Fig.~6(b) that shows the ratio $T(2\alpha_{\rm diff})/T(0)$ between the transmission of the two identical off-axis ($\alpha=2\alpha_{\rm diff}$) and on-axis ($\alpha=0$) point sources as a function of the ratio $w/r_{\rm c}$ when the coronagraph is on, taking $\Delta_\infty=\pi$, $R_1=1$~mm and $R_2=0.75R_1$ whatever $r_{\rm c}$. The apodization drawback manifests typically for $w<10r_{\rm c}$ whereas, for $w>10r_{\rm c}$, the asymptotic high-contrast regime is reached. The cutoff ratio for $w/r_{\rm c}$, of course, decreases with the separation between the sources. Finally, note that present self-engineered liquid crystal spin-orbit vortex masks are not restricted to operate at a predetermined central wavelength since $\Delta_\infty$ is electrically tunable as is the case of their artificial self-engineered counterparts \cite{slussarenko_oe_2011}.

More broadly, the proposed nature-assisted technological approach to the  fabrication of singular phase masks goes beyond the possible use in optical vortex coronagraphy, and includes  stimulated emission depletion microscopy \cite{hell_ol_1994} and spiral phase contrast microscopy \cite{furhapter_oe_2005}. Moreover, the proposed approach is not restricted to pure phase masks and opens the way to self-engineered apodization masks that are heavily used in coronagraphy \cite{carlotti_aa_2014}, which may involve the use of either transparent or absorbing liquid crystal mesophases.
Further work is needed, in particular regarding large vortex mask size issues as usually requested in low numerical aperture telescopes.

\section*{Funding Information}
This study has been carried out with financial support from the French State, managed by the French National Research Agency (ANR) in the frame of the Investments for the future Programme IdEx Bordeaux LAPHIA (ANR-10-IDEX-03-02).

\clearpage

\end{document}